# WHY NESTEDNESS IN MUTUALISTIC NETWORKS?

Enrique Burgos[1,6], Horacio Ceva[1], Roberto P.J. Perazzo[2], Mariano Devoto[3,7], Diego Medan[3,6], Martín Zimmermann[4] and Ana María Delbue[5]

[1]Departamento de Física, Comisión Nacional de Energía Atómica, Avenida del Libertador 8250, C1429BNP Buenos Aires, Argentina

[2]Departamento de Investigación y Desarrollo, Instituto Tecnológico de Buenos Aires, Av. E. Madero 399, C1106ACD Buenos Aires, Argentina

[3]Cátedra de Botánica, Facultad de Agronomía, Universidad de Buenos Aires Av. San Martín 4453, C1417DSE, Buenos Aires, Argentina

[4]Departamento de Física, Facultad de Ciencias Exactas y Naturales, Universidad de Buenos Aires, Pab. 1 Ciudad Universitaria, Buenos Aires, Argentina

[5]Departamento de Economía, Facultad de Ciencias Sociales y Económicas de la Universidad Católica Argentina, Av. A. Moreau de Justo 1400, 1107. Buenos Aires, Argentina

[6]Consejo Nacional de Investigaciones Científicas y Técnicas, Av. Rivadavia 1917, C1033AAJ, Buenos Aires, Argentina

[7]Present address: School of Biological Sciences, University of Bristol, Woodland Road, Bristol, UK, BS8 1UG

Corresponding author: Diego Medan (diemedan@agro.uba.ar)

**Abstract**

We investigate the relationship between the nested organization of mutualistic systems and their robustness against the extinction of species. We establish that a nested pattern of contacts is the best possible one as far as robustness is concerned, but only when the least linked species have the greater probability of becoming extinct. We introduce a coefficient that provides a quantitative measure of the robustness of a mutualistic system.

# 1. Introduction

Ecological systems provide a number of valuable services to mankind. The threat of global climate change as well as widespread human direct intervention has spurred the interest on ecosystems' robustness against species extinctions (Dunne et al. 2002, Memmott et al. 2004, Estrada 2007). In addition, a sustainable management of ecosystems can only be achieved with a proper understanding of how these systems are assembled and change through ecological and evolutionary time. Considerable attention has been paid to these issues (Montoya et al. 2006) for both food (Cohen et al. 1978, Pimm 2002) and mutualistic (e.g. pollination, seed dispersal) webs (Herrera and Pellmyr 2002, Jordano et al. 2003, Bascompte et al. 2003).

In this paper we concentrate on the case of mutualistic webs, of which several characteristic features have been detected. Among the most important of these features is the existence of a nested pattern of interactions in which both generalists (species holding many interactions) and specialists (holding few interactions) tend to interact with generalists whereas specialist-to-specialist interactions are infrequent (Bascompte et al. 2003).

Another feature of interest is the shape of the degree distribution of the web's nodes, i.e. the distribution of the number of contacts of each species. This has been largely studied within the framework of complex networks (Albert and Barabàsi 1999, Barabàsi and Albert 2002). These studies have been applied to the case of mutualistic networks which also show degree distributions whose decay for large values of the degree is slower than exponential ("fat tails"), and are closely fitted by truncated power laws (Jordano et al. 2003). These two characteristic features, nestedness and "fat tails", reveal that mutualistic networks are far from being a random collection of interacting species, but rather display a high degree of internal organization.

From the point of view of the theoretical analysis of mutualistic webs, one relevant goal is to explain their global features in terms of a minimal set of microscopic interactions among their constituents. In a preceding paper (Medan et al. 2007) we made progress in this direction by proposing a highly simplified picture of the process of assembly of mutualistic networks by means of a self-organizing network model (SNM).

This model accounts both for the webs' nested structure and for their degree distribution with fat tails. Although it is highly schematic, a detailed comparison of its predictions with empirical data from several ecological systems indicates that it retains most of the salient statistical features of real networks of this type. Additionally, nestedness and power-law degree distributions are shown not to be entirely independent features, as is commonly assumed. In the limit of perfect nestedness, there is a direct relationship between the shape of the degree distribution of the network and the nested pattern of interactions (see Medan et al. 2007 for further details).

Why is nestedness so ubiquitous among mutualistic networks? One of the biological reasons suggested is that it is related to the system's tolerance to species extinctions (also called robustness, Dunne et al. 2002, Memmott et al. 2004, 2005, Jordano et al. 2006). However, the reasons for the widespread occurrence of a nested pattern of contacts, and why it is a naturally emergent property of these systems remains a matter of recent debate (see Santamaría and Rodríguez-Gironés 2007 vs. Vázquez et al. 2007 in Vázquez 2007).

The fact that the SNM is a dynamical model allowed us to quantitatively assess the above statements by studying a mutualistic network at different stages of its "assembly" process and also to consider alternative assembly strategies. We are thus in a position to explore whether an increase in its nestedness is associated to an improvement in the robustness of an ecological system. In this work we use the SNM to study the relationship between system robustness and the pattern of contacts between species in a bipartite mutualist network. We aim at providing a quantitative framework to evaluate the payoff, measured in terms of robustness of the system, of a nested pattern of contacts and a power law degree distribution with a fat tail. Additionally, we introduce a coefficient to measure the robustness of a mutualistic system.

## 2. The Self-Organizing Network Model (SNM)

The interaction pattern of a mutualistic network can be coded into an adjacency matrix in which rows and columns respectively correspond to the plant and animal species of the system and 1's and 0's in the row-column intersections represent interactions or their absence, respectively. The SNM produces this matrix as the result of different possible "assembly strategies" and at different stages of the assembly process. What follows is a brief review of this model, but the reader is referred to Medan et al. (2007) for further details.

An SNM computer simulation starts from a random adjacency matrix in which the number of plants, animals and interactions are arbitrarily fixed. Starting from this (unbiased) initial configuration the SNM gradually reallocates the 1's and the 0's of the adjacency matrix following an iterative procedure. In all the iterations the following actions are alternatively applied to the rows and columns of the adjacency matrix: first two sites of a column- (row-) species respectively having a 1 and a 0 are randomly selected, and next these are swapped according to the "assembly strategy" that has been selected. One possibility [strategy (I)] is that the degree of the new partner must be higher or equal than the one of the previous partner, i.e. the contact is changed so that the interaction takes place with a more generalist counterpart. Another possibility [strategy (II)] is just the opposite, namely that the new partner must have fewer contacts than the previous one for a swap to occur.

Throughout successive iterations of either strategy the adjacency matrix eventually reaches a perfectly ordered state in which any further swapping produces no relevant change in the organization of the system. However, the two "assembly strategies" described above yield completely different patterns of interactions.

If strategy (I) is used the adjacency matrix becomes increasingly nested as more iterations are performed. If the columns and rows of the matrix are sorted according to the number of contacts, the curve that limits the contacts asymptotically approaches the 'extinction curve' studied by Atmar and Patterson (2003; 2005), also known as 'isocline of perfect nestedness' (IPN) (Bascompte et al. 2003). More interestingly, the degree distribution also changes with the number of iterations, and a truncated power law distribution emerges. Since natural systems are not perfectly ordered, comparisons with real systems (Atmar and Patterson 1995) are made by stopping the ordering process at some intermediate number of iterations, before perfect order is reached (Medan et al. 2007).

Using strategy (II), a completely different picture emerges: all species tend to have the same number of contacts and nestedness tends to disappear (Almeida-Neto et al. 2007). While strategy (I) yields networks with nestedness and degree distribution similar to those of real systems, strategy (II) fails to produce realistic results. Nevertheless, we will also consider the results obtained with the latter strategy because it provides a useful benchmark for comparison.

As an example of the use of strategy (I), Fig. 1 shows the results of running the SNM for an adjacency matrix having the same number of plants, animals and interactions as the mutualistic system studied by Clements and Long (1923). We show this matrix at different stages of the ordering process; for comparison we also show the empirically observed matrix.

## 3. Robustness

### 3.1 *The robustness coefficient*

Memmott et al (2004) gauge the robustness of a mutualist system through a curve that can be called Attack Tolerance Curve (ATC). This curve is based on the fact that if a given fraction of species of one guild (for instance, the pollinators) are eliminated (species are "attacked") a number of species of the other guild (e.g. plants) which depend on their interactions become extinct.

The ATC can be constructed from the adjacency matrix of the system by calculating the fraction of, say, row-species of the matrix with at least one contact left after all contacts of some fraction of column-species are set to 0 (Fig. 2). The ATC is contained in the unit square, starts at a value 1 on the *y*-axis, when no animal species are eliminated and all plant species survive. It then decreases monotonically reaching 1 on the *x*-axis, when no plant species survive because all animal species are eliminated. For the sake of concreteness we restrict our discussion to the case in which it is the animal species that are eliminated and, as a result, a given fraction of plant species become extinct. It is clear that an equivalent discussion can be made interchanging these roles, and that the conclusions that can be drawn are equivalent.

The above graphical description can be improved by introducing a quantitative measure of robustness with a single parameter R, defined as the area under the ATC. It is intuitive that $R \to 1$ corresponds to a curve that decreases very mildly until the point at which almost all animal species are eliminated. This is consistent with a very robust system in which most of the plant species survive even if a large fraction of the animal species is eliminated. Conversely $R \to 0$ corresponds to an ATC that decreases abruptly. This is consistent with a fragile system in which even if a very small fraction of the animal species is eliminated, most of the plants lose all their interactions and become extinct. In Fig. 2 we present an example involving several ATCs.

The assessment of the tolerance of a network through the removal of nodes bears some conceptual similarity with the studies performed by Cohen et al. 2000 to determine the resilience of the Internet under the random removal of nodes. There are however important differences that render impossible the direct application of Cohen's results to the present situation. In the first place those studies involve networks with nodes of a single type and our present analysis concerns bipartite networks. Second, the resilience of the networks such as Internet is determined by calculating the minimum fraction of nodes that have to be removed to separate the whole network into two disconnected graphs. The calculation of the ATC requires instead the removal of *an increasing fraction* of nodes of *one kind* and the subsequent determination of the number of nodes of *the other* guild that are left without counterparts and can therefore be assumed to become extinct. Consequently the definition of $R$ can not be related in any way to the segmentation of the bipartite network into disconnected graphs.

The derivation of the ATC from the adjacency matrix requires some additional assumptions. In the first place to obtain the ATC in the fashion described above no assumptions have been made concerning the "strength" of the interactions between mutualist species, i.e. how important each species of animal is for the survival of the plants with which they have contacts. This is because the extinction of a plant species is assumed only after *all* its contacts have been removed. In the second place one has to specify the order in which the animal species are eliminated.

The least biased assumption is to assume that all species have the same probability of becoming extinct. In agreement with Memmott et al. (2004) we call this the *null model*. There are two alternative, highly schematic ways in which the species can be eliminated starting either from the most connected to the least connected $(+ \to -)$ or the opposite one, from the least to the most connected $(- \to +)$. These two schemes are equivalent to assuming that species have a different probability of becoming extinct depending upon their number of contacts. The $(- \to +)$ order implies that species with less contacts with their mutualist counterparts have a greater probability of becoming extinct while the $(+ \to -)$ corresponds to the opposite.

Finally, to build the ATC it is necessary to produce a statistically significant result by averaging the calculations over several realizations. In each realization all the contacts of the same fraction of randomly selected column-species are set to 0 and then the row-species that become extinct is counted.

In the next two sections we derive the ATCs that are associated to perfectly ordered systems with each of the two assembly strategies described in section 2.

**3.2 *Robustness with assembly strategy (II)***

As mentioned above, if the SNM is run using assembly strategy (II), a perfectly ordered state is reached in which all species have the same number of contacts. This peculiar ordering is in fact quite close to a random adjacency matrix, because in this case the number of contacts in all rows fluctuates around the average. Since in this limit all animals have the same number of contacts, the null model and both the $(+ \to -)$ and $(- \to +)$ schemes provide the same ATC. We can consider this situation analytically. Let $p\,(0 \leq p \leq m)$ label the plant species in the rows, and $a\,(0 \leq a \leq n)$ label the animal species in the columns of the adjacency matrix. The probability of contacts between mutualists is $\phi$, therefore the adjacency matrix has $nm\phi$ 1's, and, asymptotically each row has $n\phi$ 1's. If a fraction $f_a = a/n$ of animal species is eliminated and $a < n\phi$, then no plant species disappear. However, if $a > n\phi$, the probability of causing an extinction of a plant species is equal to the probability $P(n,a,\phi)$ of choosing all the contacts in its row (i.e. the $n\phi$ 1's of the row). The fraction of plants that survive therefore is

$$S_p(f_a) = 1 - P(n,a,\phi) = 1 - \frac{\binom{a}{n\phi}}{\binom{n}{n\phi}} = 1 - \frac{a!(n-n\phi)!}{n!(a-n\phi)!} \qquad (1)$$

for $a > n\phi$, and $S_p(f_a) = 1$ otherwise.

### 3.3 *Robustness for perfect nestedness*

A perfectly nested system is one in which all contacts between species appear within the area delimited by the IPN. The temperature parameter $T$ calculated with the "Nestedness Calculator" (Atmar and Patterson 1995) has been proposed as a measure of the departure of a system from perfect nestedness. By convention $T=0$ corresponds to the limit in which there are no contacts outside the region delimited by the IPN. This curve has been used to analyze the nestedness of a large number of mutualistic systems (Bascompte et al. 2003), and the analytic derivation of the curve and some of its properties are discussed in Medan et al. (2007).

Let the IPN be the function $p(a)$. If animal species are ordered by decreasing number of contacts (decreasing degree) from left to right, the IPN decreases monotonously. We assume this situation in all that follows. We have defined the ATC function $S_p(f_a) = S_p(a/n)$ as the one yielding the surviving fraction of plant species when the fraction $f_a = a/n$ of animal species is eliminated. One can readily see that under the elimination scheme $(- \rightarrow +)$, $S_p(f_a)$ is a discontinuous curve: $S_p(f_a) = 1$ if $f_a < 1$, and $S_p(f_a) = 0$ if $f_a = 1$ (i.e., $a = n$). This indicates that all plant species survive the extinction of animal species until the last and most connected animal species is eliminated. A fully nested system therefore has the highest possible robustness $R^{-+} = 1$.

If, under the $(+ \rightarrow -)$ elimination scheme, all animal species between 0 and $a$ are eliminated, only those plant species $p'$ with $p(a) \leq p' \leq 1$ will become extinct. Thus, when the first $a$ species of animals are eliminated, the fraction of plant species that survive is $S_p = p(a+1)/m$. We then see that the curve $S_p(f_a)$ is nothing but the IPN normalized to the unit square. This has an important consequence because the corresponding robustness $R^{+-}$ can readily be calculated in terms of the area under the IPN, i.e., of the total number of contacts, normalized by dividing it by the product $nm$. Remarkably, the robustness associated to a fully nested system under the $(+ \rightarrow -)$ elimination strategy is therefore $R^{+-} = \phi$.

The calculation of $S_p$ for the null model is somewhat more complicated but can still be performed analytically. Assume that a number $a$ of randomly chosen animal species are eliminated and let *k+1* label the surviving animal species that has the greatest number of contacts. Since the system is perfectly nested, the fraction of plant species that survive is determined by the single *p(k + 1)*-th plant species, namely *p(k + 1)/m*. This is because the remaining animal species that are eliminated cause little harm to the bulk of the interactions between plants and animals of the system. The fraction $S_p(a/n)$ of plants that are eliminated must therefore be calculated as an average value involving $P_a(k)$ that is defined as the probability that, within the randomly chosen set of *a* columns, all columns from 1 to *k* are included in the set and the *k + 1*-th column is left out of it. We therefore have:

$$S_p(f_a) = \frac{1}{m}\sum_{k=0}^{k=a} P_a(k) p(k+1) \qquad (2)$$

The probability $P_a(k)$ is

$$P_a(k) = \frac{\binom{n-1-k}{a-k}}{\binom{n}{a}} \qquad (3)$$

as follows, for instance by a straightforward generalization of the simple cases *k = 0* and *k= 1*.

## 4. Discussion

Several ATCs derived with the two possible assembly strategies and the three elimination schemes, are compared in Fig. 2. The parameters *n*, *m* and $\phi$ involved in the corresponding equations and numerical calculations are those of the Clements and Long (1923) system.

The upper continuous thick full line corresponds to the use of strategy (II) for the IPN (Eq. (1)). In this assembly strategy there is no difference between the three possible elimination schemes. Significant differences appear instead between the two elimination schemes and the null model if the assembly strategy (I) is used. A hypothetically perfectly nested Clements and Long (1923) system gives rise to three different ATCs: the uppermost dotted horizontal line corresponds to the $(-\rightarrow+)$ elimination scheme, the lowermost, dashed curve corresponds instead to the $(+\rightarrow-)$ elimination scheme and the open circles, and intermediate curve corresponds to the null model. In this latter case we compare this analytic result using Eqs. (2) and (3) with that obtained numerically (filled triangles) from a great number (5 x$10^5$) iterations of the SNM.

The above extreme, theoretical cases can be compared in Fig. 2 to the true ATC derived from the observed data collected by Clements and Long (1923) that is not a perfectly nested system. The relevance of the SNM is shown through a remarkably good fit to the observed data obtained with 7500 iteration steps of the SNM and strategy (I).

The Fig. 2 strongly shows that the maximum robustness is achieved, for a perfectly nested system, under the $(-\rightarrow+)$ elimination scheme. However, perfected nestedness is also the most fragile under the $(+\rightarrow-)$ elimination scheme. All other situations fall between these two curves. These refer to real systems that are not perfectly nested, to both elimination schemes for strategy (II), and to the null model for strategy (I). From these results, it follows naturally that robustness critically depends upon the hypotheses that are implicit in the chosen elimination scheme and the assembly strategy.

Notice that all these conclusions coincide with those of Memmott et al. (2004). In addition we compare the ATCs for systems that are perfectly ordered under the different possible assembly strategies. From this comparison follows the remarkable result that, for the null model, strategy (II) yield ordered systems that are more robust than those generated by strategy (I).

While in figure 2 we compare observed data with extreme perfectly ordered systems, in Fig. 3 we show how the area under the ATCs changes with the number of iterations of the SNM, under each of the two assembly strategies and the three elimination schemes. In order to show all results in the same graph, curves for the assembly strategies (I) and (II) were plotted respectively in the graph's right and left panels. Iteration number 0 corresponds to a random initial adjacency matrix. A greater value of R corresponds to systems that are increasingly ordered. As can be seen from the graph the only situation in which robustness grows is for an increasing nestedness and the elimination scheme $(-\rightarrow +)$. Also notice that for a large number of iterations and the elimination scheme $(-\rightarrow +)$, $R\rightarrow 1$, while for the scheme $(+\rightarrow -)$, $R\rightarrow\phi$, as was derived in section 3.3. According to these results, nestedness happens to be the "best" organization strategy as regards robustness, *provided that species with fewer contacts have a greater probability of becoming extinct than those with more contacts.*

## 5. Conclusions

The robustness of mutualistic systems has been investigated previously by Memmott et al. (2004) by means of the ATCs, which show the fraction of species of one guild that would become extinct as a consequence of the elimination of a given fraction of species of the other guild. Extreme curves of this type can be obtained under the assumption that the attack is such that species with a greater number of contacts are eliminated either first or last. These assumptions are meant to emphasize a possible relationship between the extinction probability of a species and its number of contacts and integrate this relationship into the mathematical model.

Additionally, we investigate the dependence of *R* and the ATCs on the internal organization of the bipartite graph associated to the mutualistic system. We extend the results of Memmott et al. (2004) by using the SNM to generate a family of increasingly nested networks. We also derive analytically limiting ATCs under different hypotheses for the organization of the system. We consider situations with extreme values of nestedness, and even though these limiting contact patterns are not found in nature, they represent valuable theoretical bounds within which any real mutualistic system must fall.

By exploring the relationship between robustness and nestedness we establish a close positive relationship between the two, but *only if it is also assumed that species with less links have a greater probability of becoming extinct than those with more links.*

Conversely, if it is accepted that the only systems that are observed in nature are those with a greater robustness, the widespread occurrence of nested mutualistic systems could be regarded as an observable feature that implies the weakness of poorly linked species.

Of course, ecological specialization *per se* is not the sole factor for considering a species at greater risk of extinction; rarity, for instance, is a feature commonly linked to ecological specialization which can also contribute to the vulnerability of species to extinction (Davies et al. 2004). We are also fully aware that in this respect, other elements have been ignored such as the strength of the interaction between species (Wooton 1997, 2005). We are nevertheless confident that the ATC as calculated in the present analysis as well as the single robustness parameter *R*, can provide a good approach to the study of resiliency of the system as a whole.

We thank Dr. J. Memmott for her comments on an earlier version of this work.

**Captions to figures**

**Fig. 1**: Numerical simulation of Clements and Long's (1923) plant-pollinator system using the self-organizing network model (SNM) (Medan et al. 2007). The contacts between mutualistic species are shown as black dots. Panel (A): random initial adjacency matrix used as input to the SNM. Panel (B): adjacency matrix after 1000 iteration steps of the SNM (Atmar and Patterson 1995's parameter *T = 5.51*). Panel (C): adjacency matrix after 2175 steps, in which *T* has the same value (*T = 2.41*) than the original Clements and Long (1923) adjacency matrix shown in Panel (D). Panel (E): adjacency matrix after 100,000 steps of the SNM that corresponds to perfect nestedness (*T = 0*).

**Fig. 2**: Attack Tolerance Curves of Clements and Long's (1923) system under different elimination schemes. Dotted line: perfectly nested system under the $(-\rightarrow+)$ elimination scheme; thick continuous line: analytic result from Eq. 1 (strategy II); filled squares: after 7500 iteration steps of the SNM; empty stars (superimposed with previous curve): results for Clements and Long's (1923) original system; open circles: analytic results from Eqs. 2 and 3 (null model applied to the IPN); filled triangles: after 100,000 iteration steps of the SNM (average over 10 randomly generated realizations); dashed line: perfectly nested system under the $(+\rightarrow-)$ elimination scheme.

**Fig. 3**: Robustness *R* of the numerically simulated Clements and Long's (1923) adjacency matrix, as a function of the number of iteration steps of the SNM, and under the three kinds of elimination schemes and both assembly strategies. To ease comparison, graphs for both assembly strategies are shown together. Iteration step 0 corresponds to the random initial adjacency matrix. The number of iterations grows to the right in the right panel (strategy (I)) and to the left in the left panel (strategy (II)). Filled squares: random elimination (null model); filled stars: $(-\rightarrow+)$ elimination scheme; filled triangles: $(+\rightarrow-)$ elimination scheme. Notice that for strategy (II) all three elimination schemes converge to the same result, while for strategy (I) they tend to clearly different results. The inset shows the asymptotic behavior for strategy (I): $R \rightarrow 1$ for the $(-\rightarrow+)$ scheme, while $R \rightarrow \phi = 0.035$ for the $(+\rightarrow-)$ elimination scheme.

FIGURE 1

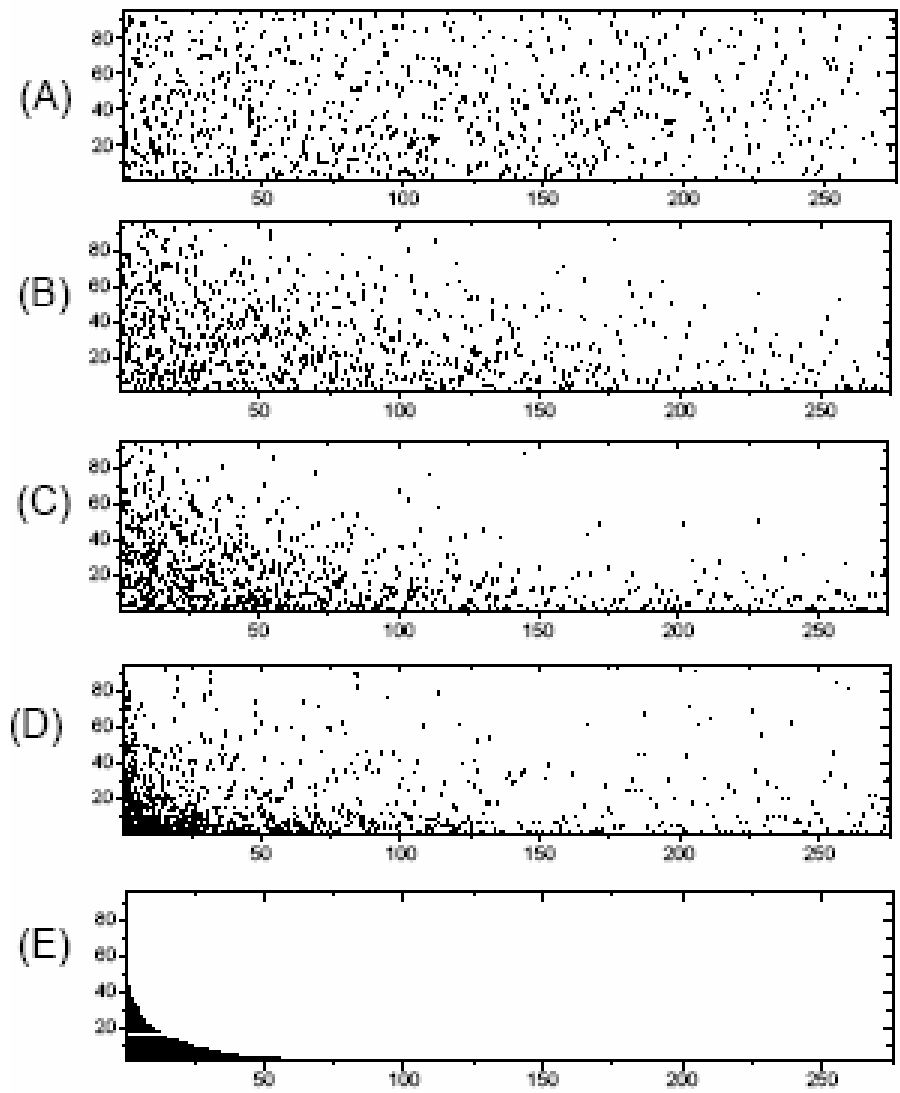

FIGURE 2

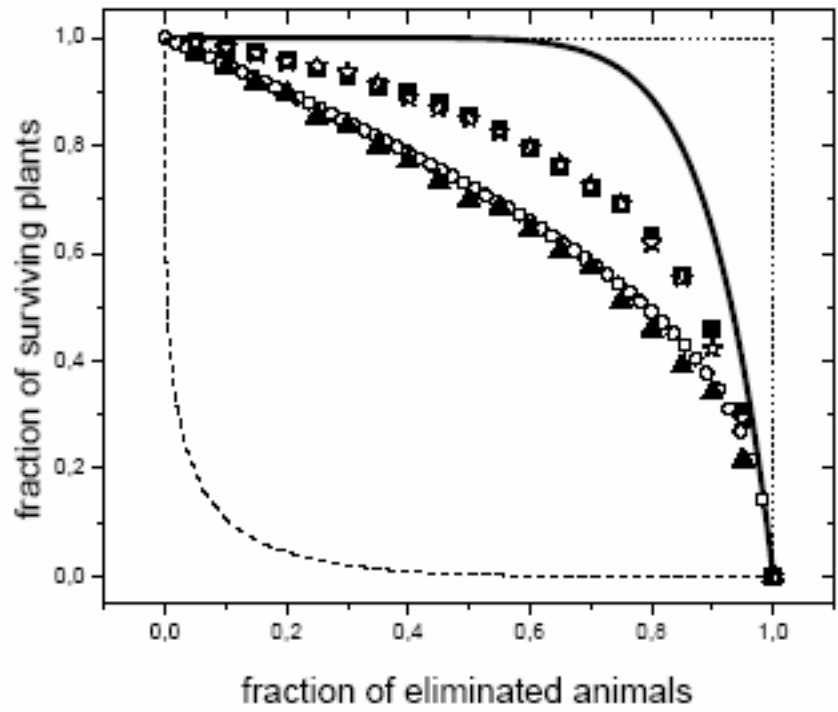

FIGURE 3

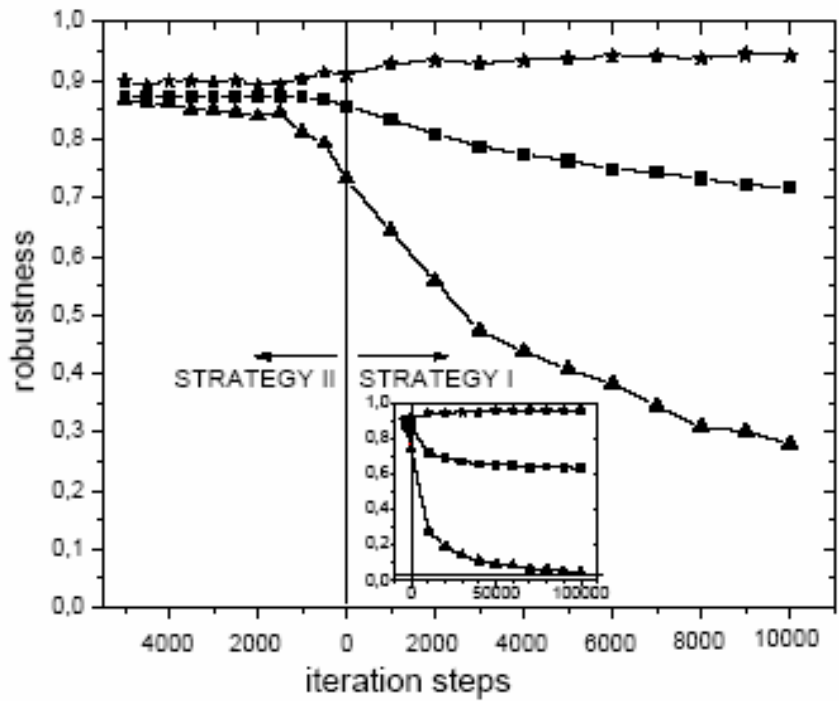